# The Infinite Square Well Is Subtle

Chyi-Lung Lin

Department of Physics, Soochow University,
Taipei, Taiwan, R.O.C.

## ABSTRACT

We show that it needs a more delicate potential to confine particles inside a well. The original model containing a vague notation $\infty$ in the potential energy is ambiguous. Using the Heaviside step function $\theta(x)$ and the Dirac delta-function $\delta(x)$, we give a precise form for the confining potential. Although such form appears unusual, the ambiguities are resolved. This form also shows that the infinite square well is not the limit of a finite square well.

**Introduction**

The infinite square well (ISW) is a model using infinitely high potential to confine particles inside a well. The notation ∞ is used in the potential energy. We also have other models using infinitely large potential energy, such as the Dirac delta-function potential. Recently, Belloni and Robinett have given a review on these models [1].

The potential energy $V(x)$ of the ISW is described by

$$V(x) = 0, \quad 0 < x < L,$$
$$= \infty, \quad \text{otherwise.} \quad (1)$$

Eq. (1) has been used for a long time, up to now. It seems dangerous in physics using a quantity such as infinity. Yet, there appear no arguments about the delta-function potential, because somehow we know how to handle the infinity; we have the formula as $\int_{-\infty}^{\infty} \delta(x) dx = 1$. On the contrary, the infinity used in the ISW is without a specification. The quantity ∞ in Eq. (1) is quite a vague notation, we have no guide on how to deal with it. As a consequence, we encounter problems in regions where $V(x) = \infty$. This includes the region outside the well and also the edges of the well. In truth, there are ambiguities resulted from straightforwardly using the notation ∞. The purpose of this paper is to show that to describe such a particle-in-a-box system needs a more delicate potential.

We first state three ambiguities in the ISW model. Two of those reside in the time independent Schrödinger equation.

$$-\frac{\hbar^2}{2m}\Psi''(x) + V(x)\Psi(x) = E\Psi(x). \quad (2)$$

The eigenfunctions $\Psi_n(x)$ and the eigenenergies $E_n$ of Eq. (2) are well-known [2-8]:

$$\Psi_n(x) = \sqrt{\frac{2}{L}} \, Sin[k_n x], \quad 0 < x < L,$$
$$= 0, \quad \text{otherwise,} \quad (3)$$

where $k_n = \frac{n\pi x}{L}$ and $n = 1, 2, 3$ .... The corresponding eigenenergy is

$E_n = \frac{\hbar^2 k_n^2}{2m}$.

The first ambiguity concerns about the query: In Eq. (2), what is the value of $V(x)\Psi(x)$ outside the well? In this region, $V(x) = \infty$ and $\Psi(x) = 0$, we inevitably encounter the ambiguity as $V(x)\Psi(x) = \infty \times 0 = ?$ From Eq. (2), the value of $V(x)\Psi(x)$ can be inferred from the values of $\Psi''(x)$ and $E\Psi(x)$. Outside the well, we have $\Psi(x) = 0$, and therefore $\Psi''(x) = 0$ and $E\Psi(x) = 0$. In order to satisfy the Schrödinger equation, it is then imposed that $V(x)\Psi(x) = 0$ outside the well. However, such a result should not be an assumption or an imposed boundary condition; it should be derived from a more precise calculation.

At the two sides of the well, Eq. (3) shows that $\Psi'(x)$ is discontinuous there. Therefore, $\Psi''(x)$ should contain a Dirac delta-function at these points. From Eq. (2), this implies that $V(x)\Psi(x)$ should as well contain delta-functions there. However, this is obscure as viewed from Eq. (1).

The second ambiguity concerns about: Can $V(x)$ bring about the discontinuity of $\Psi'(x)$ at $x = 0$ and $x = L$? Let $\Delta\Psi'(x)$ represent the amount of discontinuity of $\Psi'(x)$. Then $\Delta\Psi'(x) = \lim_{\epsilon \to 0} [\Psi'(x + \epsilon) - \Psi'(x - \epsilon)] = \lim_{\epsilon \to 0} \int_{x-\epsilon}^{x+\epsilon} \Psi''(y)\, dy$. From Eq. (2), we can substitute $\Psi''(x)$ by $V(x)\Psi(x)$ and $E\Psi(x)$. Using the result that $\lim_{\epsilon \to 0} \int_{x-\epsilon}^{x+\epsilon} E\Psi(y)\, dy \to 0$, we have

$$\Delta\Psi'(x) = \frac{2m}{\hbar^2} \lim_{\epsilon \to 0} \int_{x-\epsilon}^{x+\epsilon} V(y)\Psi(y)\, dy. \qquad (4)$$

The integral in the right side of Eq. (4) determines the value of $\Delta\Psi'(x)$ [9]. If $\Delta\Psi'(x) = 0$, then $\Psi'(x)$ is continuous at $x$, and vice versa. However, we cannot calculate $\Delta\Psi'(0)$ and $\Delta\Psi'(L)$ directly from Eq. (4). We note that $V(x)\Psi(x) = 0$ inside the well, this is because $V(x) = 0$ in this region. And we also have $V(x)\Psi(x) = 0$ outside the well. Together, we have $V(x)\Psi(x) = 0$ both inside and outside the well. From Eq. (1), it is unclear about the values of $V(x)\Psi(x)$ at the two sides. The integral in Eq. (4) therefore is ambiguous. On the other hand, from Eq. (3), we have $\Delta \Psi_n{}'(0) = \sqrt{\frac{2}{L}}\, k_n$ and $\Delta \Psi_n{}'(L) = -\sqrt{\frac{2}{L}}\, k_n Cos(k_n L)$. These two results mean that $V(x)\, \Psi_n(x)$ must contain a Dirac delta-function at $x = 0$ and $x = L$ [10]. Hence, to have the solution of $\Psi_n(x)$ in Eq. (3), it requires a more delicate potential energy than that given in Eq. (1).

The third ambiguity concerns about the manifestation of the Ehrenfest's theorem. The ambiguity lies in that we cannot calculate the expectation value $\left\langle \Psi(t) \left| \left(-\frac{\partial V(x)}{\partial x}\right) \right| \Psi(t) \right\rangle$; it concerns an integration of $x$ over the range $[0, L]$. Yet, the term $\frac{dV(x)}{dx}$, which is related to force, is zero inside the well, but is infinitely large at the boundaries. While, the probability density, $|\Psi(x,t)|^2$, is zero at the boundaries. Hence we again encounter the ambiguity as $\infty \times 0 = ?$ [11].

All of these three ambiguities are resulted from the lack of a precise form for the infinity contained in $V(x)$. As we cannot tackle the infinity, some results are then obtained by imposition. In fact, we obtain the solution in Eq. (3) by imposing the boundary condition that $\Psi(x)$ is continuous at the sides of the well. And we also impose the condition that $V(x)\, \Psi(x) = 0$ outside the well. We should note that imposing conditions on $\Psi(x)$ and $V(x)\, \Psi(x)$ means that we are loosing the connection between $V(x)$ and $\Psi(x)$. It seems that we have put in some results by imposition which are beyond what can be derived from the $V(x)$. We might say that the potential energy in Eq. (1) does not correspond to the true potential energy for the wave function $\Psi(x)$ described in Eq. (3).

On the other hand, we may start from known solutions of $\Psi(x)$, and go back to determine the corresponding potential energy. We should therefore be able to derive the precise form of $V(x)$ from the known solutions in Eq. (3).

We first rewrite the separate solutions in Eq. (3) into one form. Using the notation of Heaviside step function $\theta(x)$, we can rewrite Eq. (3) in a more compact form as

$$\Psi_n(x) = G(x)\, \theta(x)\, \theta(L - x). \qquad (5)$$

$$G(x) = \sqrt{\frac{2}{L}}\, Sin[k_n\, x]. \qquad (6)$$

Eq. (5) may also be written as: $\Psi_n(x) = G(x)[\theta(x) - \theta(L-x)]$[1, 12]. Similarly, we may try to give a form for the $V(x)$ in terms of $\theta(x)$ as

$$V(x) = \frac{1}{\theta(x)} + \frac{1}{\theta(L-x)} - 2. \tag{7}$$

Then we have $V(x) = \infty$ outside the well, and $V(x) = 0$ inside the well. The $V(x)$ in Eq. (7) seems to represent a potential energy for the ISW. However, from Eqs. (5) and (7), we obtain

$$V(x)\,\Psi(x) = G(x)\,[\theta(x) + \theta(L-x) - 2\theta(x)\,\theta(L-x)\,]. \tag{8}$$

This yields $V(x)\,\Psi(x) = G(x)$ outside the well. As we need $V(x)\,\Psi(x) = 0$ in this region, this means that $G(x) = 0$. We conclude that the potential energy in Eq. (7) only leads to a trivial solution which is $\Psi(x) = 0$ everywhere. Thus ISW is subtle. It requires that the function $V(x)$ should be carefully determined. We need a more precise and accurate form of $V(x)$ for nontrivial solutions of $\Psi(x)$.

## $V(x)$ derived from the known solutions of ISW

We rewrite Eq. (2) as

$$\frac{\hbar^2}{2m}\Psi''(x) + E\,\Psi(x) = V(x)\,\Psi(x). \tag{9}$$

Substituting the $\Psi_n(x)$ in Eq. (3) into the left side of (9), we can then read out the form of $V(x)$. Using the following results of $\theta(x)$ and the Dirac delta-function $\delta(x)$:

$$\theta'(x) = \delta(x), \tag{10}$$
$$f(x)\,\delta'(x-a) = -f'(a)\,\delta(x-a), \tag{11}$$

we obtain:

$$\Psi_n'(x) = G'(x)\,\theta(x)\,\theta(L-x) + G(x)\,\delta(x)\,\theta(L-x)$$
$$-G(x)\,\theta(x)\,\delta(L-x). \tag{12}$$

$$\Psi_n''(x) = G''(x)\,\theta(x)\,\theta(L-x) + G'(x)\,\delta(x)\,\theta(L-x)$$

$$-G'(x)\,\theta(x)\,\delta(L-x) \qquad (13)$$

Substituting Eqs. (5), (6), and (13) into Eq. (9), we obtain

$$V(x)\,\Psi_n(x) = \sqrt{\frac{2}{L}}\frac{\hbar^2}{2m}[\,k_n\,\delta(x)\,\theta(L-x) - k_n\,Cos(k_n L)\,\theta(x)\,\delta(L-x)\,]. \qquad (14)$$

The first term of the square bracket in Eq. (14) can be rewritten as

$$k_n\,\delta(x)\,\theta(L-x)$$

$$= \frac{k_n\,\delta(x)}{Sin(k_n x)\,\theta(x)}\,\Psi_n(x)$$

$$= \frac{\delta(x)}{x\,\theta(x)}\,\Psi_n(x). \qquad (15)$$

Above we have used $\frac{\delta(x)}{Sin(k_n x)} \to \frac{\delta(x)}{k_n\,x}$. To rewrite the second term, we use the following result

$$Sin(k_n x) = Sin(k_n(x-L))\,Cos(k_n L)\,. \qquad (16)$$

Using Eq. (16), the second term can be rewritten as

$$-k_n\,Cos(k_n L)\,\theta(x)\,\delta(L-x)$$

$$= -\frac{k_n\,Cos(k_n L)\,\delta(L-x)}{Sin(k_n x)\,\theta(L-x)}\,\Psi_n(x)$$

$$= -\frac{k_n\,\delta(L-x)}{Sin(k_n(x-L))\,\theta(L-x)}\,\Psi_n(x)$$

$$= \frac{\delta(L-x)}{(L-x)\,\theta(L-x)}\,\Psi_n(x). \qquad (17)$$

Substituting Eqs. (15), and (17) into Eq. (14), we obtain

$$V(x)\,\Psi_n(x) = \frac{\hbar^2}{2m}\left[\frac{\delta(x)}{x\,\theta(x)} + \frac{\delta(L-x)}{(L-x)\,\theta(L-x)}\right]\Psi_n(x). \quad (18)$$

This yields the precise form of the potential energy for the ISW:

$$V(x) = \frac{\hbar^2}{2m}\left[\frac{\delta(x)}{x\,\theta(x)} + \frac{\delta(L-x)}{(L-x)\,\theta(L-x)}\right]. \quad (19)$$

We can show that from the potential energy $V(x)$ in Eq. (19), the solutions of Eq. (2) are those given by Eq. (3). In what follows we show that the previous three ambiguities can be resolved with the potential energy $V(x)$ given in Eq. (19).

**About ambiguity (1): The value of $V(x)\,\Psi_n(x)$ outside the well**

We note that the $V(x)$ in Eq. (20) is quite subtle. It is similar to the $V(x)$ in Eq. (7), but there are more subtle descriptions at the two sides of the well. From Eq. (19), we in fact cannot obtain an explicit value of $V(x)$ outside the well, which is of the form as $\frac{zero}{zero}$. Hence, the potential energy outside the well is represented by an unmanageable form. However, the product $V(x)\,\Psi_n(x)$ in this region is explicit. From Eqs. (5), (6) and (19), we have

$$\begin{aligned}V(x)\,\Psi_n(x)\\ = \frac{\hbar^2}{2m}\left[\frac{\delta(x)}{x\,\theta(x)} + \frac{\delta(L-x)}{(L-x)\,\theta(L-x)}\right]\sqrt{\frac{2}{L}}\,Sin(k_n\,x)\,\theta(x)\,\theta(L-x)\\ = \sqrt{\frac{2}{L}}\frac{\hbar^2}{2m}\{\,k_n\,\delta(x) - k_n\,Cos(k_n\,L)\,\delta(L-x)\,\}.\end{aligned} \quad (20)$$

From Eq. (20), we have $V(x)\,\Psi_n(x) = 0$ outside the well. It should be noted that this result is a derived-result. It is not an imposed boundary condition. We also note that $V(x)\,\Psi_n(x)$ indeed contains delta-functions at the sides of the well. These two singularities will then contribute to the discontinuity of $\Psi'(x)$ at the two sides.

**About ambiguity (2): Can $V(x)$ bring about the discontinuity of $\Psi'(x)$ at $x = 0$ and $x = L$?**

From Eqs. (4), and (21), we can now directly calculate $\Delta\Psi_n{'}(x)$.

We have

$$\Delta\Psi_n'(0) = \frac{2m}{\hbar^2}\int_{-\epsilon}^{\epsilon} V(x)\,\Psi_n(x)\,dx$$

$$= \sqrt{\frac{2}{L}}\,k_n \int_{-\epsilon}^{\epsilon} \delta(x)\,dx.$$

$$= \sqrt{\frac{2}{L}}\,k_n, \qquad (21)$$

and

$$\Delta\Psi_n'(L) = \frac{2m}{\hbar^2}\int_{L-\epsilon}^{L+\epsilon} V(x)\,\Psi_n(x)\,dx$$

$$= -\,k_n\,Cos(k_n L)\int_{-\epsilon}^{\epsilon} \delta(x-L)\,dx.$$

$$= -\,k_n\,Cos(k_n L). \qquad (22)$$

Eqs. (21) and (23) show that the product $V(x)\Psi_n(x)$ does offer contributions to the discontinuity of $\Psi_n'(x)$ at $x = 0$ and $x = L$. The obtained values of $\Delta\Psi_n'(0)$ and $\Delta\Psi_n'(L)$ are consistent with those calculated from Eq. (3).

**About ambiguity (3): The Ehrenfest's theorem**

We now discuss the manifestation of the Ehrenfest's theorem for time-evolved wave packets in the ISW. The time evolution of a general wave packet $\Psi(x,t)$ is as follows

$$\Psi(x,t) = \sum_{n=1}^{\infty} a_n\,\Psi_n(x)\,e^{-i\omega_n t}, \qquad (23)$$

where $\omega_n = \frac{E_n}{\hbar}$. To verify the Ehrenfest's theorem for packets, we need to verify the following formula:

$$\frac{d}{dt}\langle\Psi(t)|\,P\,|\Psi(t)\rangle = -\langle\Psi(t)|\frac{dV(x)}{dx}|\Psi(t)\rangle, \qquad (24)$$

where $P$ represents the momentum operator. Since we have the explicit form of $V(x)$ in Eq. (19), we can calculate the expectation values directly. The expectation value is calculable, as we have

$$\frac{dV(x)}{dx} = \frac{\hbar^2}{2m} \left[ \frac{\delta'(x)}{x\,\theta(x)} - \frac{\delta(x)}{x^2\theta(x)} - \frac{\delta(x)^2}{x\,\theta(x)^2} + \frac{\delta'(x-L)}{(L-x)\,\theta(L-x)} \right.$$
$$\left. + \frac{\delta(x-L)}{(L-x)^2\theta(L-x)} + \frac{\delta(L-x)^2}{(L-x)\,\theta(L-x)^2} \right]. \qquad (25)$$

Multiplying Eq. (25) by $\Psi_n(x)\,\Psi_m(x)$, the step functions in the denominator are cancelled out. Each term in $\Psi_n(x)\frac{dV(x)}{dx}\Psi_m(x)$ is then regular, and the integration of which over the range $[0, L]$, is then finite. We whence obtain the final result

$$\left\langle \Psi(t) \left| \frac{dV(x)}{dx} \right| \Psi(t) \right\rangle$$
$$= \int_{-\infty}^{\infty} \Psi^*(x,t)\, \frac{dV(x)}{dx}\, \Psi(x,t)\, dx$$
$$= -\frac{\hbar^2}{m\,L} \sum_{n=1}^{\infty} \sum_{\substack{m=1 \\ (m \neq n)}}^{\infty} a_n^*\, a_m\, k_n\, k_m\, \beta_{nm}\, e^{i(\omega_n - \omega_m)t}, \qquad (26)$$

where $\beta_{nm} = 1 - (-1)^{n+m}$. Together with the following results

$$\langle \Psi(t) | P | \Psi(t) \rangle$$
$$= (-i\,\hbar)\,\frac{2}{L} \sum_{n=1}^{\infty} \sum_{\substack{m=1 \\ (m \neq n)}}^{\infty} a_n^*\, a_m\, \frac{k_n\, k_m}{(k_n^2 - k_m^2)}\, \beta_{nm}\, e^{i(\omega_n - \omega_m)t}. \qquad (27)$$

And

$$\frac{d}{dt} \langle \Psi(t) | P | \Psi(t) \rangle$$
$$= \frac{\hbar^2}{m\,L} \sum_{n=1}^{\infty} \sum_{\substack{m=1 \\ (m \neq n)}}^{\infty} a_n^*\, a_m\, k_n\, k_m\, \beta_{nm}\, e^{i(\omega_n - \omega_m)t}. \qquad (28)$$

Comparing Eqs. (26) and (28), we see that the Ehrenfest's theorem is confirmed for arbitrary time-evolved packets in the ISW. The calculations are done directly using the function $V(x)$ in Eq. (19). We need not first do those calculations in the finite square well and then take the limit [11]. It has been argued that the ISW is not the limit of a finite well [13]. Our form of $V(x)$ in Eq. (19) supports this argument.

**Conclusion**

We discuss a quantum system in which particles are confined to a finite region of space. The original ISW model uses infinitely large potential to forbid particles going outside. The potential energy contains the vague notation ∞. However, constructing confining potential is subtle. It needs a more delicate potential to confine particles inside a well. The behavior of the confined particles is described in Eq. (3). We use this solution and the time independent Schrödinger equation to derive the corresponding potential. The potential energy $V(x)$ derived is expressed in terms of the step function $\theta(x)$ and the Dirac delta-function $\delta(x)$. The infinity ∞ originally appeared outside the well and at the two sides is replaced by $\theta(x)$ and $\delta(x)$. In other words, the infinity is now with a specification and can be managed. The form of the function $V(x)$ appears unusual, yet, this just reflects that to confine particles inside a well is not that straightforward and is uncommon.

This way of constructing confining potential may also be used to other similar models.

# Acknowledgement

The author would like to thank Prof. Young-Sea Huang and Prof. Tsin-Fu Jiang for many helpful discussions and suggestions.